\documentclass[final]{svjour3}
\usepackage{graphicx}
\usepackage{rotating}
\usepackage{amssymb}
\usepackage{mathptmx}
\usepackage[numbers]{natbib}
\makeatletter
\journalname{Journal of Low Temperature Physics}
\usepackage{tabularx}
\usepackage{subcaption}
\usepackage{wrapfig}

\bibpunct{}{}{,}{s}{}{,}

\begin{document}

\newcommand{\hdblarrow}{H\makebox[0.9ex][l]{$\downdownarrows$}-}
\title{Performance of a phonon-mediated detector using KIDs optimized for sub-GeV dark matter}

\author{O. Wen, T. Aralis, R. Basu Thakur, B. Bumble, Y.-Y. Chang, K. Ramanathan, S.R. Golwala}

\institute{Division of Physics, Mathematics, \& Astronomy, California Institute of Technology,\\ Pasadena, CA 91106, USA
\email{owwen@caltech.edu}}

\maketitle

\begin{abstract}

Detection of sub-GeV dark matter candidates requires sub-keV detector thresholds on deposited energy. We provide an update on a gram-scale phonon-mediated KID-based device that was designed for a dark matter search in this mass range at the Northwestern Experimental Underground Site. Currently, the device is demonstrating 6 eV resolution on the energy absorbed by the resonator. With some important assumptions, this translates to 20 eV baseline resolution on energy deposited in the substrate. We show that TLS noise dominates this energy resolution estimate. After modifying the design to mitigate TLS noise, we project 5 eV baseline resolution on energy deposited in the substrate (1.5 eV on energy absorbed by the resonator) for an amplifier-white-noise-dominated device. Finally, we present a clear path forward to sub-eV thresholds, which includes installation of a quantum-limited superconducting parametric amplifier and adjustments to the material makeup of our resonators.


\end{abstract}

\section{Objective: Low mass dark matter searches}

There is a shifting paradigm in the dark matter community to focus its efforts on particle masses outside of the WIMP mass range from 1GeV to 1TeV [1]. The low mass dark matter range (LDM) from 1keV to 1GeV is an interesting mass range for near-term experiments because of the rich population of theory targets and the ease with which existing experiments can adapt to this mass range. 1MeV dark matter mass corresponds to 1eV energy thresholds in particle detectors, requiring sub-eV resolutions on energy deposits. Since these energies are below electronic excitation levels, phonon detectors are naturally suited for probing this range. A TES-based phonon detector from the SuperCDMS collaboration has demonstrated 3.86eV resolution [2]. Kinetic inductance detectors (KIDs) are already competitive with this resolution and are well-suited for bringing sub-eV energy resolutions to phonon detectors. 

\begin{figure}[htbp]
\begin{center}
\includegraphics[width=0.8\linewidth, keepaspectratio]{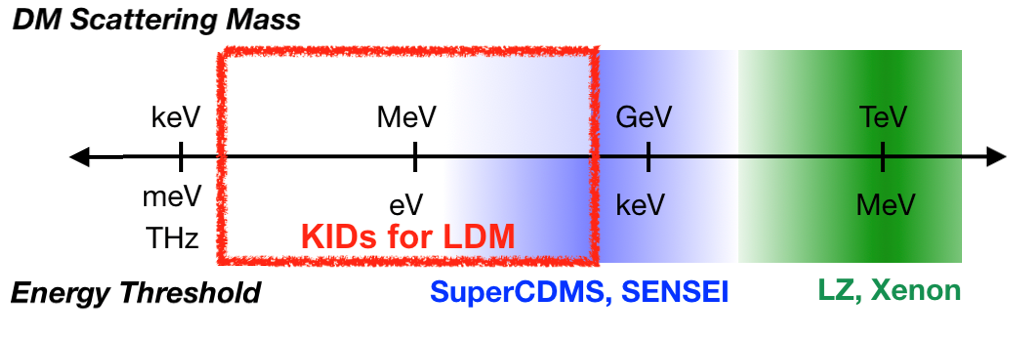}
\caption{Energy range of interest for kinetic inductance detectors, compared against existing dark matter direct detection experiments. (Color figure online.)}
\end{center}
\label{LDM}
\end{figure}

\section{KID design: optimizing for lower energy thresholds}

There are two major design choices that were made to fabricate a device optimized for lower energy thresholds:

First, there is only a single resonator designed to be the phonon collecting resonator. The baseline energy resolution of a KID-based device scales with $\sqrt{\textrm{\# of KIDs}}$, so to minimize threshold, we design for a single phonon collecting resonator in the middle of the device. This resonator is made out of aluminum. There are 10 other resonators on the device, which are used for calibration purposes.

Second, we minimize the amount ``dead metal'', which is any metal on the chip that might absorb phonons and not contribute to signal. This is done by fabricating all other features on the device out of niobium. Niobium's $T_c$ is 10x greater than aluminum's, which means that a phonon must be 10x more energetic to break a Cooper pair in the niobium film. These other features include: bonding pads, the feedline, the other resonators, and even the capacitor of the phonon-collecting resonator. 

\begin{figure}[htbp]
\centering 
\begin{subfigure}{0.45\textwidth}
\centering
\includegraphics[width=\textwidth, keepaspectratio]{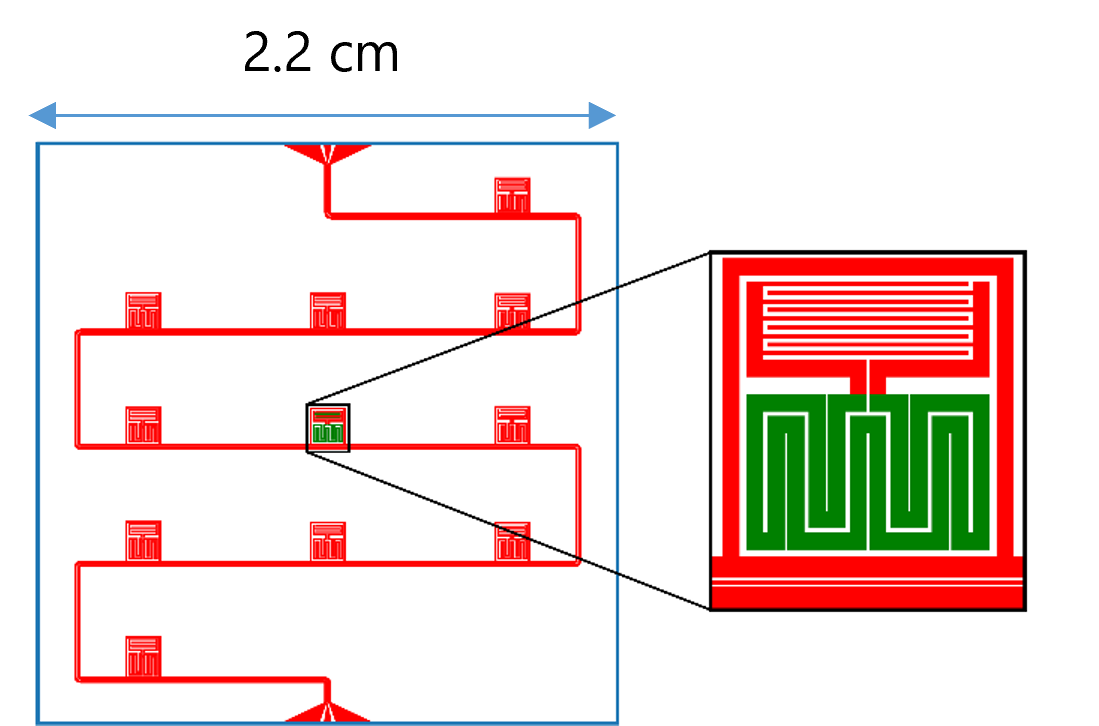}
\end{subfigure}
\begin{subfigure}{0.45\textwidth}
\centering
\includegraphics[width=\textwidth, keepaspectratio]{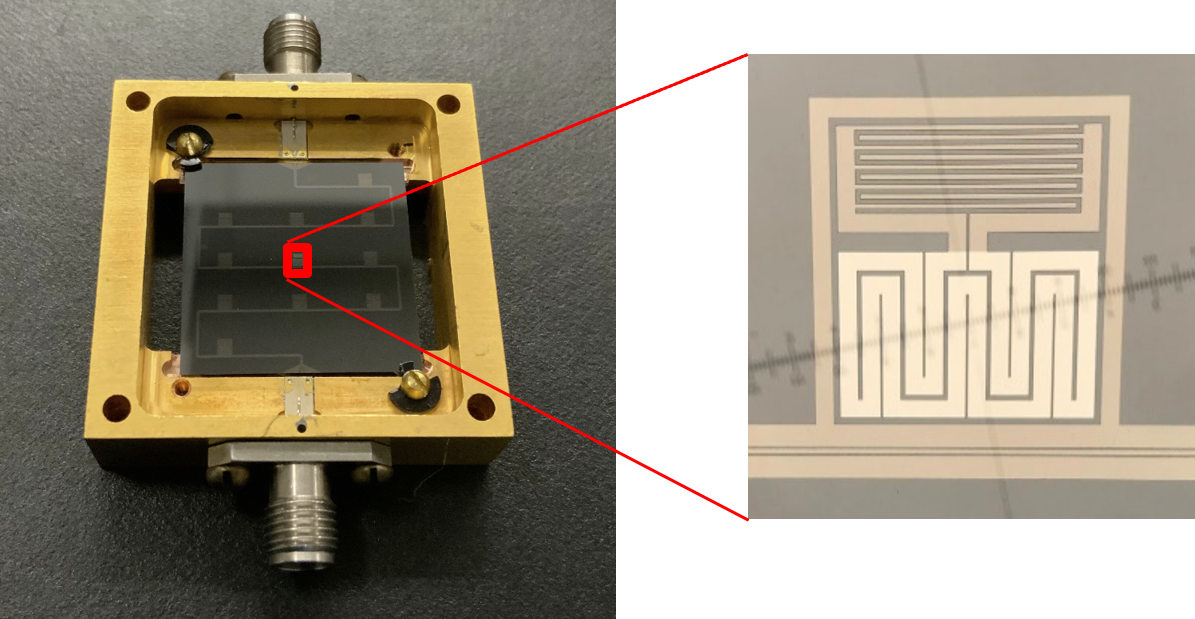}
\end{subfigure}
\caption{{\it Left}: Device design; {\it red} is niobium and {\it green} is aluminum. {\it Right}: Device in its device box. {\it Both zoom-ins}: aluminum resonator; top part is the interdigitated capacitor and the bottom is the meandering inductor. (Color figure online.)}
\label{device}
\end{figure}

\begin{figure}[htbp]
\begin{center}
\includegraphics[width=\linewidth, keepaspectratio, trim= {0 4cm 0 4cm},clip]{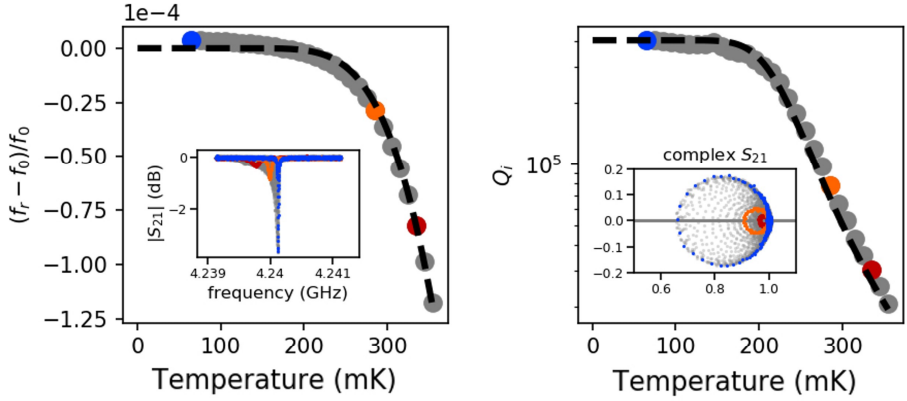}
\caption{ {\it Left}: Fractional change in resonator frequency versus temperature. {\it Blue}, {\it orange}, and {\it red} correspond to 65mK, 285mK, and 335mK. {\it Left inset}: Magnitude of $S_{21}$ at the entire range of temperatures, with three temperatures highlighted in their corresponding colors to show the shift in resonance frequency. {\it Right}: Resonator internal quality factor versus temperature. Color coding same as before. {\it Right inset}: Complex $S_{21}$ at the entire range of temperatures, with three temperatures highlighted in their corresponding colors to show the changing quality factor. Fit values are: $\Delta$ = 0.184meV; $\alpha$ = 3.801\%; $f_0$ = 4.2401GHz; and $Q_{i0}$ = 405538. (Color figure online.)}
\end{center} 
\label{mb_fits}
\end{figure}

\section{KID basics: resonator characterization}

To measure the amount of energy that gets deposited in the resonator, we track the number of quasiparticles in the resonator that result from broken Cooper pairs. The conversion of our measurement from electronics units to quasiparticle units requires the superconducting bandgap $\Delta$, which is half of the phonon energy required to break a Cooper pair, and the kinetic inductance fraction $\alpha$. There is a straightforward procedure for measuring these values:

\begin{enumerate}
\item Measure $S_{21}$ versus frequency using a vector network analyzer.
\item Fit for resonator's resonance frequency $f_r$ and quality factor $Q_r$.
\item Repeat steps 1 and 2 at a range of temperatures between $\sim$10\% of $T_c$ and $\sim$30\% of $T_c$.
\item Fit for $\Delta$ and $\alpha$ using Mattis-Bardeen equations [3].
\end{enumerate}

The measured values for $\Delta$ and $\alpha$ are close to those seen in a device of similar design.

\begin{figure}
\centering
\begin{subfigure}{0.55\textwidth}
\centering
\includegraphics[width=\textwidth]{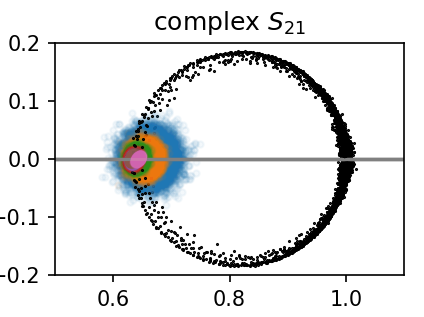}
\end{subfigure} 
\centering
\begin{subfigure}{0.3\textwidth}
\centering
\includegraphics[width=\textwidth]{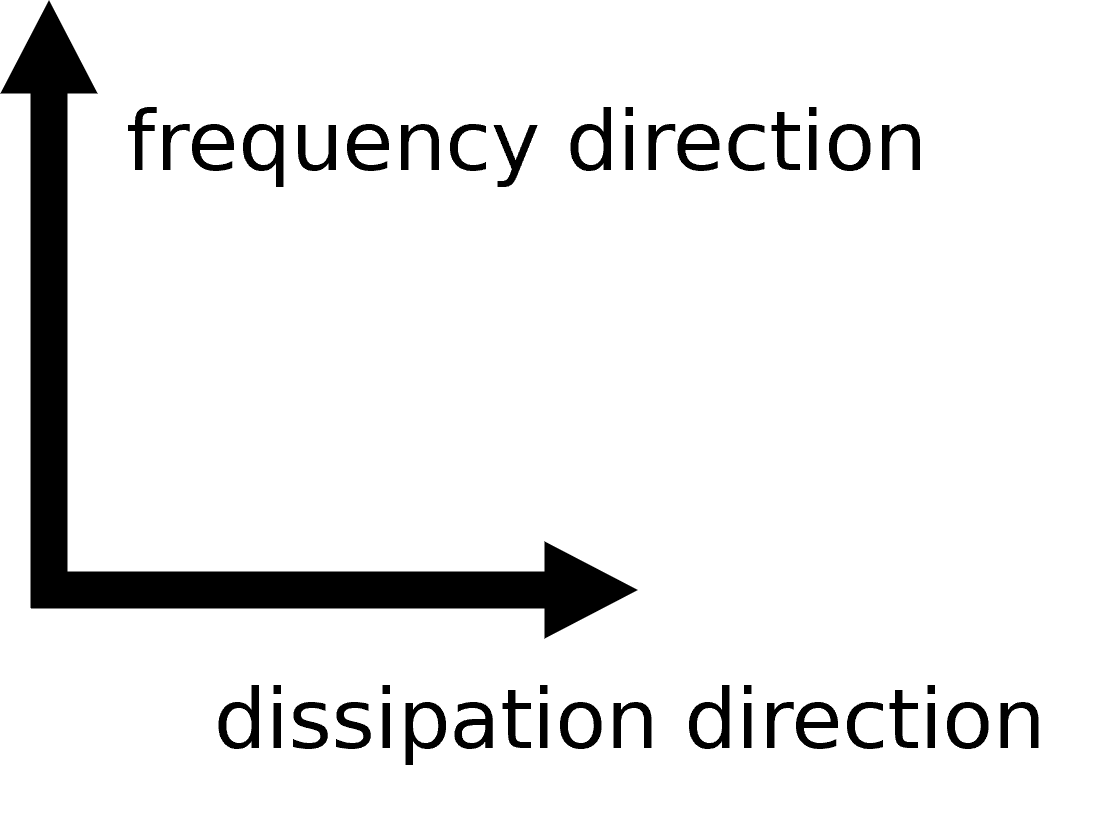}
\end{subfigure} 
\begin{subfigure}{0.45\textwidth}
\centering
\includegraphics[width=\textwidth]{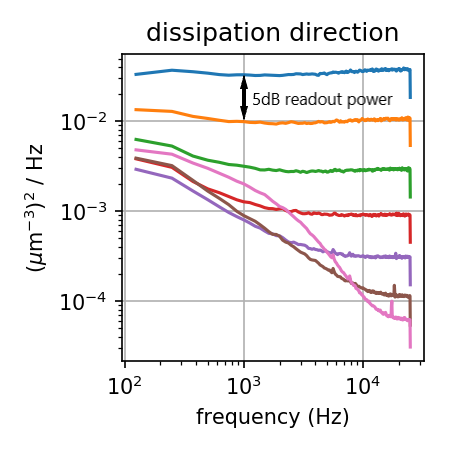}
\end{subfigure}
\begin{subfigure}{0.45\textwidth}
\centering
\includegraphics[width=\textwidth]{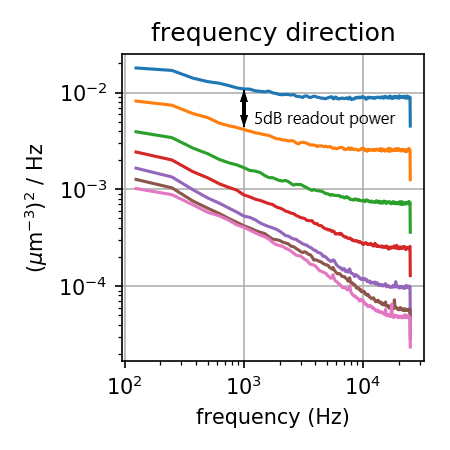}
\end{subfigure}
\caption{{\it Top}: $S_{21}$ noise traces are shown for seven different readout powers separated by 5dBm, corresponding to seven different colors, with {\it blue} being the lowest power and {\it pink} being the highest power.  {\it Black} points are VNA scans. {\it Bottom}: Power spectral densities for the frequency and dissipation directions. (Color figure online.)}
\label{k1 and k2 noise}
\end{figure}

\section{KID operation: $S_{21}$ readout and its noise}

A kinetic inductance detector is readout by measuring $S_{21}(f_r)$ as a function of t: $S_{21}(f_r; t)$. The instrument we use to perform this measurement is an Ettus Research USRP Software Defined Radio device, which uses a 200MHz bandwidth signal, that is then digitally mixed to recover $S_{21}(f_r; t)$. During data-taking, $S_{21}(f_{\mathrm{off}};t)$ of an off-resonance tone $f_{\mathrm{off}}$ is simultaneously measured in order to monitor and clean out noise in $S_{21}(f_r; t)$ that is not caused by the resonance, e.g. multiplicative gain and phase noise from the amplifier and USRP. Further low-pass filtering and decimation is done during analysis.

\begin{figure}
\centering
\begin{subfigure}{0.45\textwidth}
\centering
\includegraphics[width=\textwidth]{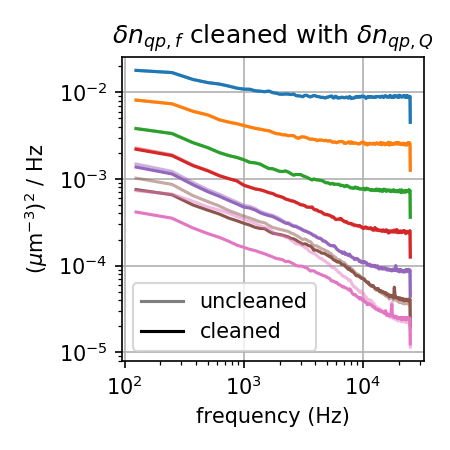}
\end{subfigure} 
\begin{subfigure}{0.45\textwidth}
\centering
\includegraphics[width=\textwidth]{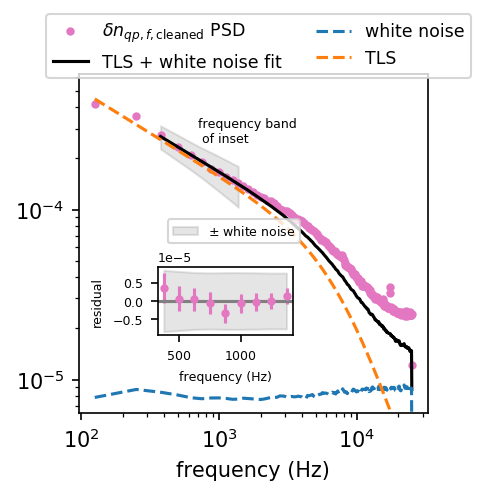}
\end{subfigure}
\begin{subfigure}{0.5\textwidth}
\centering
\includegraphics[width=\textwidth]{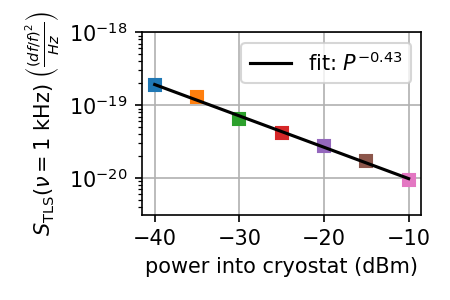}
\end{subfigure}
\caption{{\it Top left}: frequency direction PSDs before and after cleaning, i.e. removal of correlated noise with dissipation direction. {\it Top right}: cleaned frequency direction PSD of the highest readout power, fit to a model composed of TLS noise and white noise. {\it Top right inset}: residuals of the fit for a subset of the frequency range, along with bounds given by the white noise. {\it Bottom}: TLS level in $\frac{df}{f}$ units versus readout power; power law fit results also shown. (Color figure online.)}
\label{TLS figs}
\end{figure}

The $S_{21}$ versus frequency scan by the vector network analyzer is used to identify the frequency and dissipation directions for $S_{21}(f_r; t)$. This is the typical basis that is used for tracking quasiparticle production in the resonator; the two directions correspond to the directions of maximal change in $Q_r$ and $f_r$ [3]. The Mattis-Bardeen fits for $\Delta$ and $\alpha$ are used to convert $\delta S_{21}(f_r, t)$ into quasiparticle densities in the resonator: $\delta n_{qp,f}(t)$ and $\delta n_{qp,Q}(t)$. Power spectral densities (PSDs) of the quasiparticle density timestreams are calculated along these two directions and are shown for seven different powers in Fig.~\ref{k1 and k2 noise}. 

In the dissipation direction at the lowest powers ({\it blue} and {\it orange}), the noise is white, and the dependence on power is as expected: a 5dB increase in readout power corresponds to a half decade decrease in the power at all frequencies. This dependence on power persists up through the second highest power ({\it brown}) at frequencies above 10 kHz. At the highest readout powers, there is an unknown non-white noise source that dominates at frequencies below 10 kHz. This manifests itself as a tilt in the noise trace that is most presently seen at the highest power ({\it pink}).

In the frequency direction, the shapes of the power spectral density curves resemble two-level system (TLS) noise [4], but the tilted noise trace is an indication that the unknown noise in the dissipation direction is also present in the frequency direction. Thus, we use the dissipation direction to clean the frequency direction; specifically, this means removal of the correlated components between $\delta n_{qp,Q}(t)$ and $\delta n_{qp,f}(t)$ from $\delta n_{qp,f}(t)$ via the following computation:
\begin{equation}
\delta n_{qp,f,\mathrm{cleaned}}(t) = \delta n_{qp,f}(t) - A_{Q,f}\delta n_{qp,Q}(t), \quad\mathrm{ where }\quad A_{Q,f} = \frac{\mathrm{Cov}(\delta n_{qp,Q},\delta n_{qp,f})}{\mathrm{Var}(\delta n_{qp,f})}
\end{equation}

After the correlated noise is removed from $\delta n_{qp,f}(t)$ and the noise PSDs are recalculated (Fig.~\ref{TLS figs} {\it Top left}), the cleaned noise PSDs are fit to a model that consists of white noise and TLS noise (Fig.~\ref{TLS figs} {\it Top right}). In a frequency band that is dominated by TLS noise, which coincides with the frequency band of the signal template, the fit agrees with the data to a level that is smaller than the white noise (Fig.~\ref{TLS figs} {\it Top right inset}). Furthermore, when the TLS level at 1kHz is plotted versus readout power, we find the dependence on readout power to be $P^{-0.43}$ (Fig.~\ref{TLS figs} {\it Bottom}), which is close to the usual power dependence of TLS: $P^{-0.5}$ [4].

\section{Calculating energy resolutions}

The baseline energy resolution on quasiparticle density can be calculated from the following result of the optimal filter formalism [5]:
\begin{equation}
\sigma^2_A = \left[T\sum^{\frac{N}{2}-1}_{n=-\frac{N}{2}}\frac{|\tilde{s}_n|^2}{J(f_n)}\right]^{-1}
\end{equation}
where A is taken to be the quasiparticle density in units of $\mu$m$^{-3}$, $T$ is the total duration of the timestream, $N$ is the number of frequency bins, $\tilde{s}_n$ is the Fourier coefficients of the signal template, and $J(f_n)$ is the noise power spectral density.

In the provided calculations, we set $J(f_n)$ as the $\delta n_{qp,f,\mathrm{cleaned}}$ PSD. To produce the signal template, we use a novel technique: we drive one of the niobium resonators at its resonance frequency with a large amount of readout power; this breaks Cooper pairs in the resonator that then recombine and send phonons into the substrate. The phonons are then absorbed by the aluminum readout resonator, and a pulse shape is formed.

Importantly, we note that $\sigma_A$ is in units of quasiparticle density. This can be translated into $\sigma_{E,\mathrm{res}}$, the resolution on the energy that is absorbed by the resonator, using $\Delta$ and the resonator's geometry. To compute $\sigma_E$, the resolution on the energy deposited in the substrate, we must assume some phonon collection efficiency $\eta_{ph}$. We assume a 30\% collection efficiency, based off literature values [6]. We emphasize that this is the largest source of uncertainty in our measurement of $\sigma_E$

\begin{wrapfigure}{r}{0.5\textwidth}
\vspace{-1em}
\centering
\textbf{Table of energy resolutions} \\
\begin{tabular}{ | m{2em} | m{6em} | m{6em} | }
 \hline
   & TLS-limited device & amplifier-limited device \\ 
 \hline
  $\sigma_{E,\mathrm{res}}$ &  6 eV  &  1.5 eV \\  
 \hline
  $\sigma_{E}$ & 20 eV  & 5 eV \\  
 \hline
\end{tabular}
\vspace{-1em}
\end{wrapfigure}

Finally, we present a hypothesis on the TLS-limited noise that we see in this device: since we have not seen TLS noise in a similarly designed aluminum only device [7] and we also see TLS noise in the accompanying niobium resonators, the TLS noise we observe is caused by the niobium cap on the aluminum resonator's capacitor. Thus, we propose a new device that replaces the niobium layer on the capacitor with stochiometric TiNx, a material that has demonstrated low TLS noise. This material can be tuned to have a $T_c$ in between that of aluminum and niobium, so that we need not degrade our phonon collection efficiency. We can project the noise performance of such a device by using our dissipation direction measurement of the white noise at the lowest readout power and propagating the white noise to its expected level at the highest readout power. 

\section{Future work: next steps toward sub-eV energy thresholds}

The immediate next step is to perform an absolute energy calibration using an LED laser. This has already been installed into the cryostat and testing is already underway [8]. Next, once we have demonstrated amplifier-limited noise on our $S_{21}$ measurement, we can improve that noise by a factor of 5 through use of a kinetic inductance parametric amplifier. Lastly, there are also plans to replace the aluminum film with a lower $T_c$ material such as aluminum manganese, which may provide up to a factor of 10 improvement in resolution.

\begin{acknowledgements}
We acknowledge the support of the following institutions and grants: NASA, NSTGRO 80NSSC20K1223 Department of Energy, DE-SC0011925F Fermilab, LDRD Subcontract 672112
\end{acknowledgements}

\pagebreak

\end{document}